\documentclass[twocolumn,showpacs,preprintnumbers,amsmath,amssymb]{revtex4}

\usepackage{graphicx}
\usepackage{dcolumn}
\usepackage{bm}

\begin{document}


\title{First results of the search for neutrinoless double beta decay with the NEMO~3 detector}

\author{R.~Arnold$^e$, C.~Augier$^h$, J.~Baker$^d$, A.~Barabash$^f$, G.~Broudin$^a$, V.~Brudanin$^g$, A.J.~Caffrey$^d$, E.~Caurier$^e$, V.~Egorov$^g$, K.~Errahmane$^h$, A.I.~Etienvre$^h$, 
J.L.~Guyonnet$^e$, F.~Hubert$^a$, Ph.~Hubert$^a$, C.~Jollet$^e$, S.~Jullian$^h$, O.~Kochetov$^g$, V.~Kovalenko$^g$, S.~Konovalov$^f$, D.~Lalanne$^h$, F.~Leccia$^a$, C.~Longuemare$^i$, G.~Lutter$^a$, Ch.~Marquet$^a$, F.~Mauger$^i$, F.~Nowacki$^e$ H.~Ohsumi$^l$, F.~Piquemal$^a$, J.L.~Reyss$^j$, R.~Saakyan$^m$, X.~Sarazin$^h$, 
L.~Simard$^h$, F.~\v{S}imkovic$^o$, Yu.~Shitov$^g$, A.~Smolnikov$^g$, I.~\v{S}tekl$^c$, J.~Suhonen$^n$, C.S.~Sutton$^k$, G.~Szklarz$^h$, J.~Thomas$^m$, V.~Timkin$^g$, V.~Tretyak$^g$, V.~Umatov$^f$, L.~V\'ala$^c$, I.~Vanushin$^f$, V.~Vasilyev$^f$, V.~Vorobel$^b$}
\author{Ts.~Vylov$^g$}

\affiliation{
$^a$CENBG, IN2P3-CNRS and UBI, 33170 Gradignan, France\\
$^b$Charles University, Prague, Czech Republic\\
$^c$IEAP, Czech Tech. Univ., Prague, Czech Republic\\
$^d$INL, Idaho Falls, ID 83415, U.S.A.\\
$^e$IReS, IN2P3-CNRS and ULP, 67037 Strasbourg, France\\
$^f$ITEP, 117259 Moscow, Russia\\
$^g$JINR, 141980 Dubna, Russia\\
$^h$LAL, IN2P3-CNRS and UPS, 91405 Orsay, France\\
$^i$LPC, IN2P3-CNRS and UC, 14032 Caen, France\\
$^j$LSCE, CNRS, 91190 Gif-sur-Yvette, France\\
$^k$MHC, South Hadley, Massachusetts 01075, U.S.A.\\
$^l$Saga University, Saga 840-8502, Japan\\
$^m$UCL, London WC1E 6BT, United Kingdom\\
$^n$Jyv\"askyl\"a University, FIN-40351 Jyv\"askyl\"a, Finlande\\
$^o$FMFI,~Comenius~Univ.,~SK-842~48~Bratislava,~Slovakia}

\date{\today}

\begin{abstract}
The NEMO~3 detector, which has been operating in the Fr\'ejus underground laboratory since 
February 2003, is devoted to the search for neutrinoless double beta decay ($\beta\beta$0$\nu$). 
The half-lives of the two neutrino double beta decay ($\beta\beta 2 \nu$) have been measured for $^{100}$Mo and $^{82}$Se.
After 389 effective days of data collection from February 2003 until September 2004 (Phase~I), no evidence for neutrinoless double beta decay was found from $\sim$7~kg of $^{100}$Mo and $\sim$1~kg of $^{82}$Se. 
The corresponding limits are $T_{1/2}(\beta\beta0\nu)>4.6 \times 10^{23}$ years for $^{100}$Mo and $T_{1/2}(\beta\beta0\nu)>1.0 \times 10^{23}$ years for $^{82}$Se (90\% C.L.). 
Depending on the nuclear matrix element calculation, the limits for the effective Majorana neutrino mass are $\langle m_{\nu} \rangle < 0.7-2.8$~eV for $^{100}$Mo and $\langle m_{\nu} \rangle < 1.7-4.9$~eV for $^{82}$Se.
\end{abstract}


\maketitle

\section{\label{sec:level1}Introduction}

The positive results obtained in the last few years in neutrino oscillation experiments \cite{SuperKSolar,SuperKAtm,SNO,Kamland} have demonstrated that neutrinos are massive particles and that lepton flavor is not conserved. 
In parallel, tritium $\beta$-decay experiments \cite{Mainz,Troitsk} have established a very low limit on the electron neutrino mass of $m_{\nu_e} < 2.2$~eV (95\% CL).
Grand Unified Theories can provide a natural framework for neutrino masses and lepton number violation. In particular the see-saw model \cite{Mohapatra1980} which requires the existence of a Majorana neutrino, naturally explains the smallness of neutrino masses. 
The existence of  Majorana neutrinos would also provide a natural framework for the leptogenesis mechanism \cite{Leptogenesis} which could explain the observed baryon-antibaryon asymmetry in the universe.
The observation of neutrinoless double beta decay ($\beta\beta$0$\nu$) would prove that neutrinos are Majorana particles and that global lepton number is not conserved.
It would also constrain the mass spectrum and the absolute mass of the neutrinos. \\


\section{The NEMO~3 detector}

The NEMO~3 detector \cite{Augier2005}, installed in the Fr\'ejus underground laboratory (LSM, France) is  searching for $\beta\beta 0 \nu$ decay by the direct detection of the two electrons with a combination of tracking and calorimeter information. 
The two main isotopes present inside the detector in the form of very thin foils (40-60 mg/cm$^2$) are $^{100}$Mo (6914~g, $Q_{\beta\beta}=3034 keV$) and $^{82}$Se (932~g, $Q_{\beta\beta}=2295 keV$).
On both sides of the sources, there is a gaseous tracking detector which consists of $6180$ open drift cells operating in the Geiger mode allowing three-dimensional track reconstruction. 
To minimize the multiple scattering, the gas is a mixture of 95\% helium, 4\% ethyl alcohol, 1\% argon and 0.1\% water.
The wire chamber is surrounded by a calorimeter which consists of $1940$ plastic scintillator blocks coupled to very low radioactive photomultipliers (PMTs). 
The energy resolution (FWHM) of the calorimeter is 14\% at 1 MeV for the scintillators equipped with a 5'' PMTs on the external wall and 17\% for the 3'' PMTs on the internal wall.
The resolution of the summed energy of the two electrons in the $\beta\beta 0 \nu$ decay is mainly a convolution of the energy resolution of the calorimeter and the fluctuation in the electron energy loss in the foil source which gives a non-gaussian tail. The FWHM of the expected two-electron energy spectrum of the $\beta\beta 0 \nu$ decay is 350~keV.
Absolute energy calibrations are carried out every 40 days using $^{207}$Bi sources.
A daily laser survey controls the gain stability of each PMT.
A solenoid surrounding the detector produces a 25 Gauss magnetic field in order to distinguish electrons from positrons. 
An external shield of 18 cm of low radioactivity iron, a water shield and a wood shield  cover the detector to reduce external $\gamma$ and neutrons. \\

\section{Measurement of $\beta\beta 2 \nu$ decays}

\begin{figure}
\includegraphics[scale=0.25]{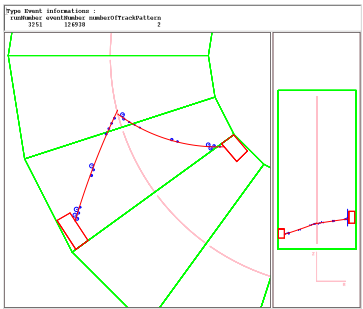}
\caption{\label{fig:event} Transverse (left) and longitudinal (right) view of a reconstructed $\beta\beta$ event selected from the data with a two electron energy sum of 2812 keV.}
\end{figure}

\begin{figure*}
\includegraphics[scale=0.4]{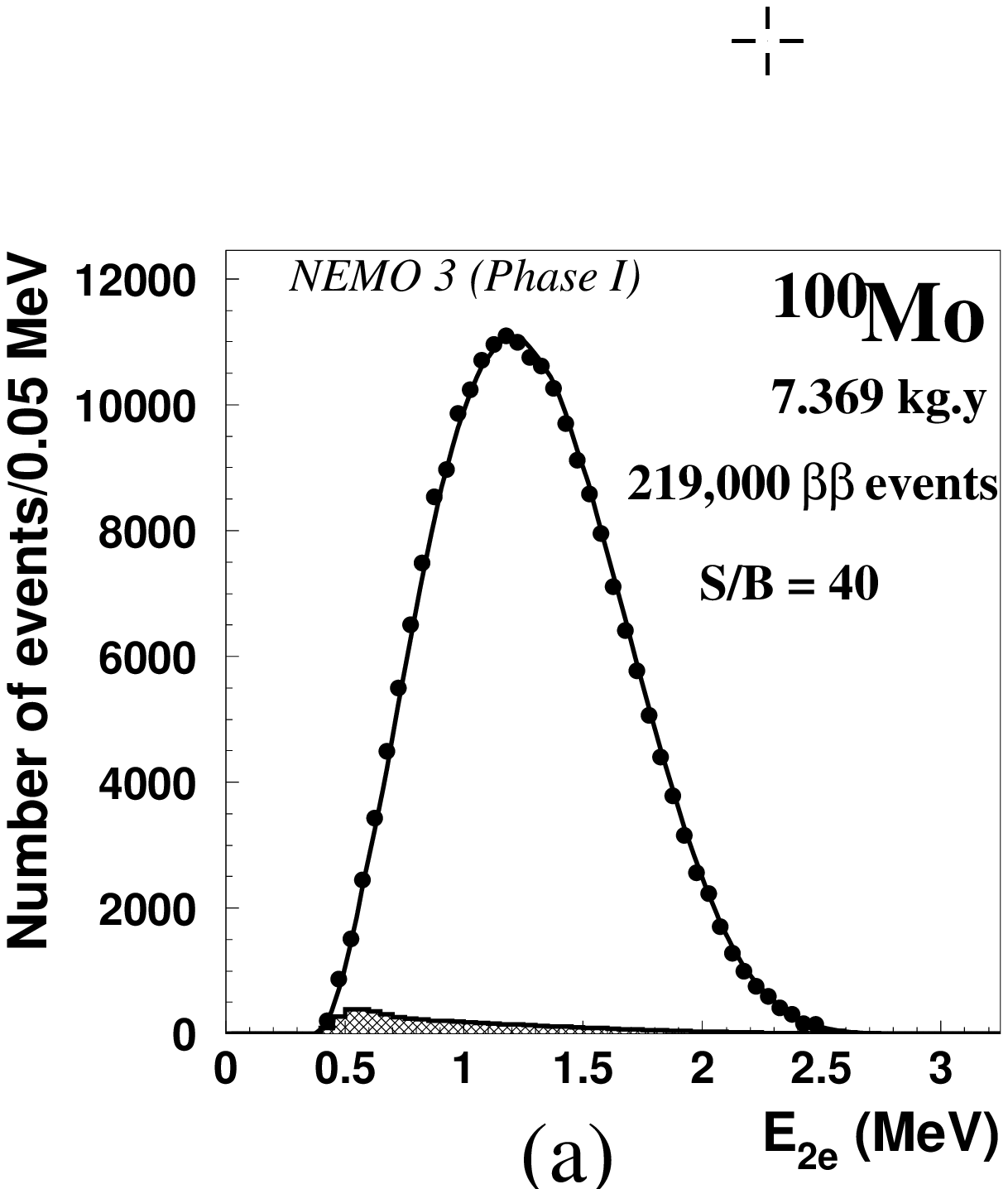}
\includegraphics[scale=0.4]{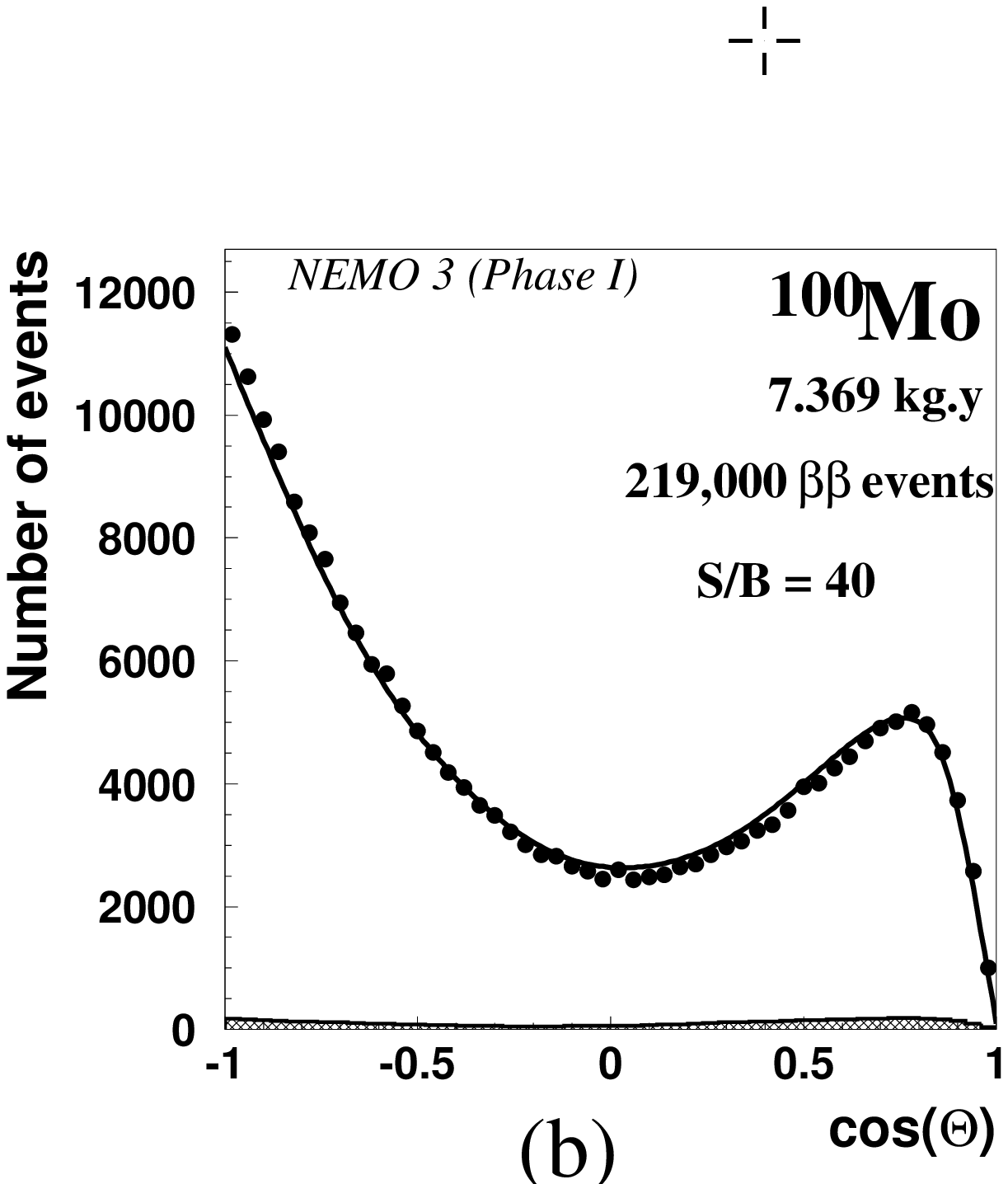}
\includegraphics[scale=0.4]{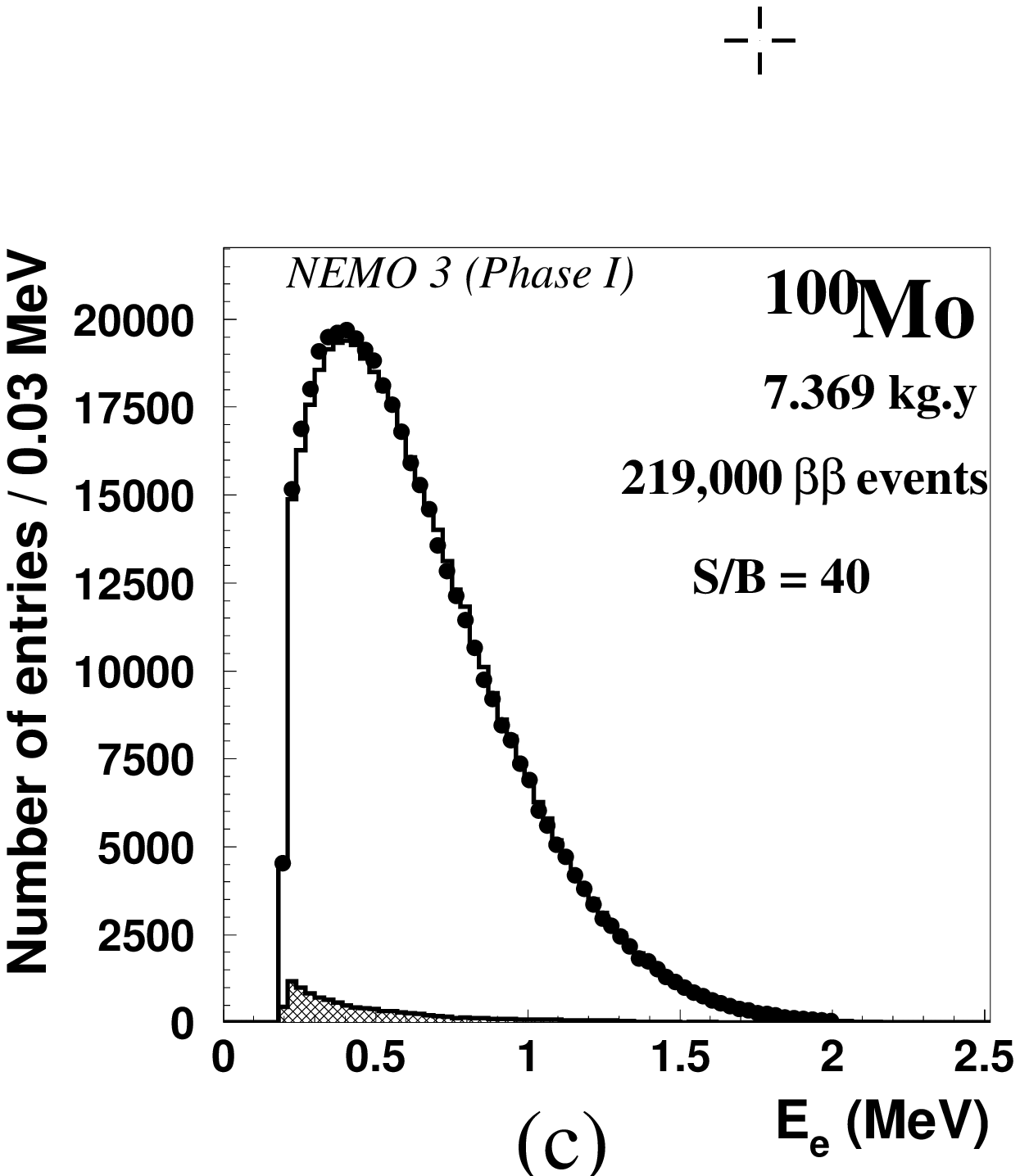}
\caption{\label{fig:mobb2nu} (a) Energy sum spectrum of the two electrons, (b) angular 
distribution of the two electrons and (c) single energy spectrum of the electrons, after background 
subtraction from $^{100}$Mo with 7.369 kg.years  exposure. The solid line corresponds to the expected 
spectrum from $\beta\beta 2 \nu$ simulations and the shaded histogram is the subtracted background 
computed by Monte-Carlo simulations. The signal contains 219,000 $\beta\beta$ events and the signal-to-background ratio is 40.}
\end{figure*}

\begin{figure}
\includegraphics[scale=0.4]{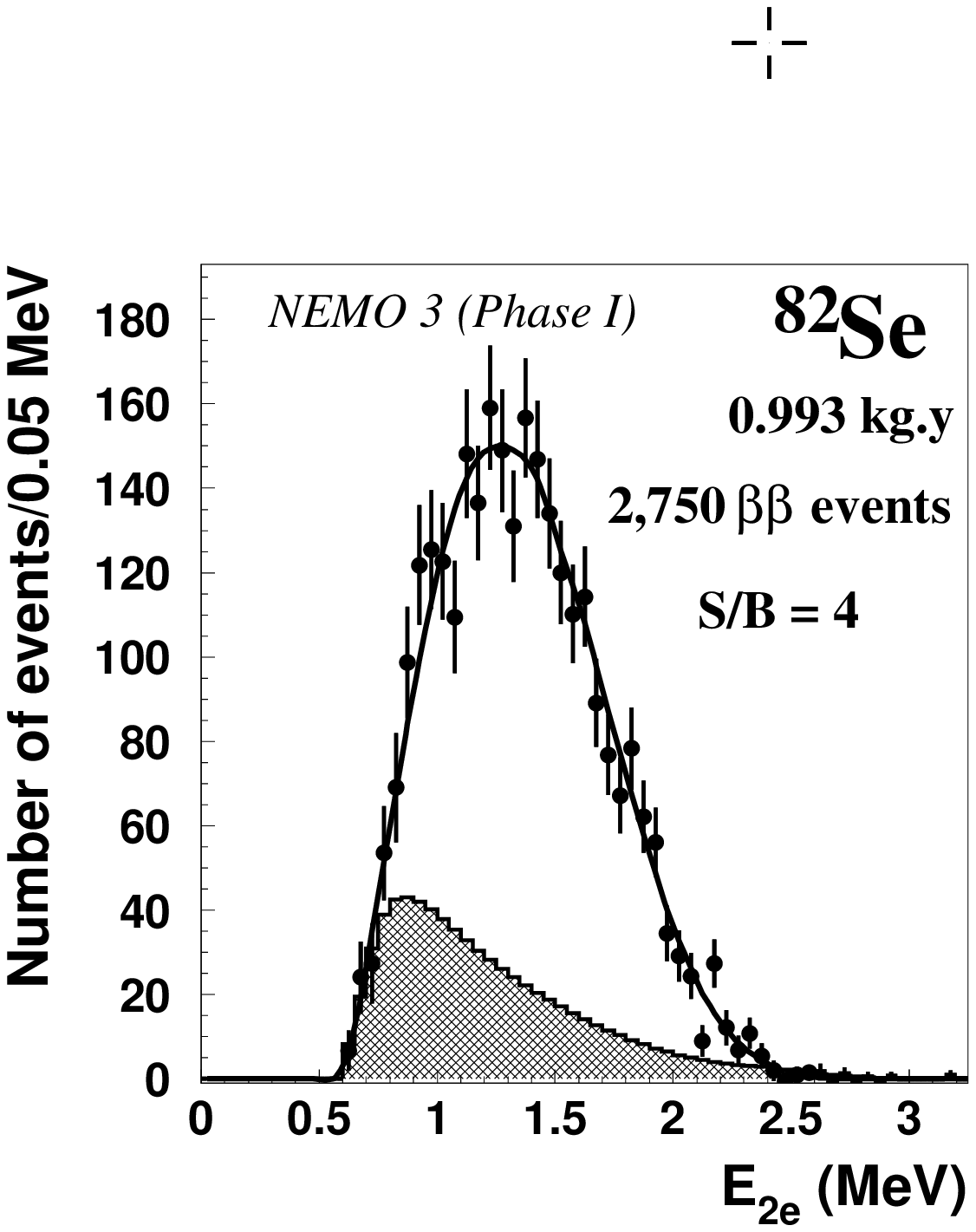}
\caption{\label{fig:sebb2nu} Energy sum spectrum of the two electrons after background subtraction 
from $^{82}$Se with 0.993 kg.years exposure (same legend as Fig.~\ref{fig:mobb2nu}). The signal contains 
 2,750 $\beta\beta$ events and the signal-to-background ratio is 4.}
\end{figure}

A two-electron (2$e^-$) event (see Fig.~\ref{fig:event}) candidate for a $\beta\beta$ decay is defined as follows: 
two tracks come from the same vertex on the source foil, each track must be associated with a fired scintillator,  
its curvature must correspond to a negative charge and the time-of-flight must correspond 
to the two electrons being emitted from the same source position. 
For each electron an energy threshold of 200~keV for $^{100}$Mo and 300~keV for $^{82}$Se is applied. 
Fig.~\ref{fig:mobb2nu}(a) and Fig.~\ref{fig:sebb2nu} show the two-electron energy sum spectra after background subtraction 
obtained after 389 effective days of data collection with $^{100}$Mo and with $^{82}$Se respectively. 
The angular distribution of the two electrons and the single energy spectrum are also 
presented in the case of $^{100}$Mo in Fig.~\ref{fig:mobb2nu}(b) and (c). 
All these spectra are in good agreement with the $\beta\beta 2 \nu$ simulations. 
The values of the measured half-lives are 
$T_{1/2}(\beta\beta2\nu) = [ 7.11 \pm 0.02(stat) \pm 0.54(syst) ] \times 10^{18}$~y for $^{100}$Mo 
(with a Single State Dominance decay) and 
$[ 9.6 \pm 0.3(stat) \pm 1.0(syst) ] \times 10^{19}$~y 
for $^{82}$Se. 
These values are in agreement with, but have a higher precision than the previous measurements\cite{PDG}.\\

\section{Study of the background in the $\beta\beta 0 \nu$ energy window}

A complete study of the background in the $\beta\beta 0 \nu$ energy window has been performed. 
The level of each background component has been directly measured using different analysis channels in the data. 

External backgrounds due to $^{214}$Bi and $^{208}$Tl contaminants outside the source foils (mostly in the PMTs) have been 
measured by searching for Compton electrons emitted from the source foils by external $\gamma$.  
For $^{208}$Tl, a total activity of $\sim$40~Bq has been measured and is in agreement with the previous HPGe measurements of 
samples of the PMT glass. For $^{214}$Bi, an activity of $\sim$300~Bq has been found, again in agreement with the HPGe 
measurements of PMTs and also the level of radon surrounding the detector inside the shield. 
The expected number of $\beta\beta 0 \nu$-like events due to this background is negligible, 
$\lesssim10^{-3}$~counts.kg$^{-1}$.y$^{-1}$ in the $[2.8-3.2]$ MeV energy window where the $\beta\beta 0 \nu$ signal is expected.

External neutrons and high energy $\gamma$ backgrounds have been measured by searching for 
crossing electron events above 4 MeV. 
This corresponds to a negligible expected level of background of $\sim 3\times10^{-3}$~counts.kg$^{-1}$.y$^{-1}$ 
in the $\beta\beta 0 \nu$ energy window.

The level of $^{208}$Tl impurities inside the sources has been measured by searching for 
internal ($e^-\gamma\gamma$) and ($e^-\gamma\gamma\gamma$) events. The measured activity is 
$80 \pm 20$~$\mu$Bq/kg in molybdenum and $300 \pm 50$~$\mu$Bq/kg in Selenium. 
It is in agreement with the previous HPGe measurements which gave an upper limit of 100~$\mu$Bq/kg for Molybdenum and a 
positive measurement of $400 \pm 100$~$\mu$Bq/kg for Selenium. 
This corresponds to an expected level of background in the $\beta\beta 0 \nu$ energy window of 
$\sim 0.1$~counts.kg$^{-1}$.y$^{-1}$ for Molybdenum and $\sim 0.3$~counts.kg$^{-1}$.y$^{-1}$ for Selenium. 
The measurement of $^{214}$Bi impurities inside the sources could not be achieved in this first period of 
data due to Radon contamination (see later). 
However the previous HPGe measurements gave an upper limit of 350~$\mu$Bq/kg for Molybdenum and a positive 
measurement of $1.2 \pm 0.5$~mBq/kg for Selenium, corresponding to a negligable expected level of backgound.

The expected level of background due to the tail of the $\beta\beta 2 \nu$ distribution in the 
$\beta\beta 0 \nu$ energy window is $\sim0.3$~counts.kg$^{-1}$.y$^{-1}$ for Molybdenum and 
$\sim0.02$~counts.kg$^{-1}$.y$^{-1}$ for Selenium.

The dominant background in this first period of data was Radon gas inside the tracking chamber due to a 
low rate of diffusion of Radon from the laboratory ($\sim$15~Bq/m$^3$) into the detector. 
Two independent measurements of the Radon level in the detector were carried out. 
The first used a high sensitivity Radon detector similar to the one developed by the Super-Kamiokande 
collaboration \cite{Takeuchi1999}. 
The second was done by searching for ($e^-$, {\it delayed-}$\alpha$) events in the NEMO~3 data. 
Indeed the tracking detector allows the detection of the delayed tracks (up to 700~$\mu$s later) in order 
to tag delayed-$\alpha$ emitted by $^{214}$Po in the Bi-Po process. 
Both measurements are in good agreement and indicate a level of radon inside the detector of $25 \pm 5$~mBq/m$^3$. 
This radon contamination corresponds to an expected level of background in the 
$\beta\beta 0 \nu$ energy window of $\sim$1~count.kg$^{-1}$.y$^{-1}$. \\

\section{First results on the limitsearch of $\beta\beta 0 \nu$ decay with $^{100}$Mo and $^{82}$Se}

\begin{figure*}
\includegraphics[scale=0.3]{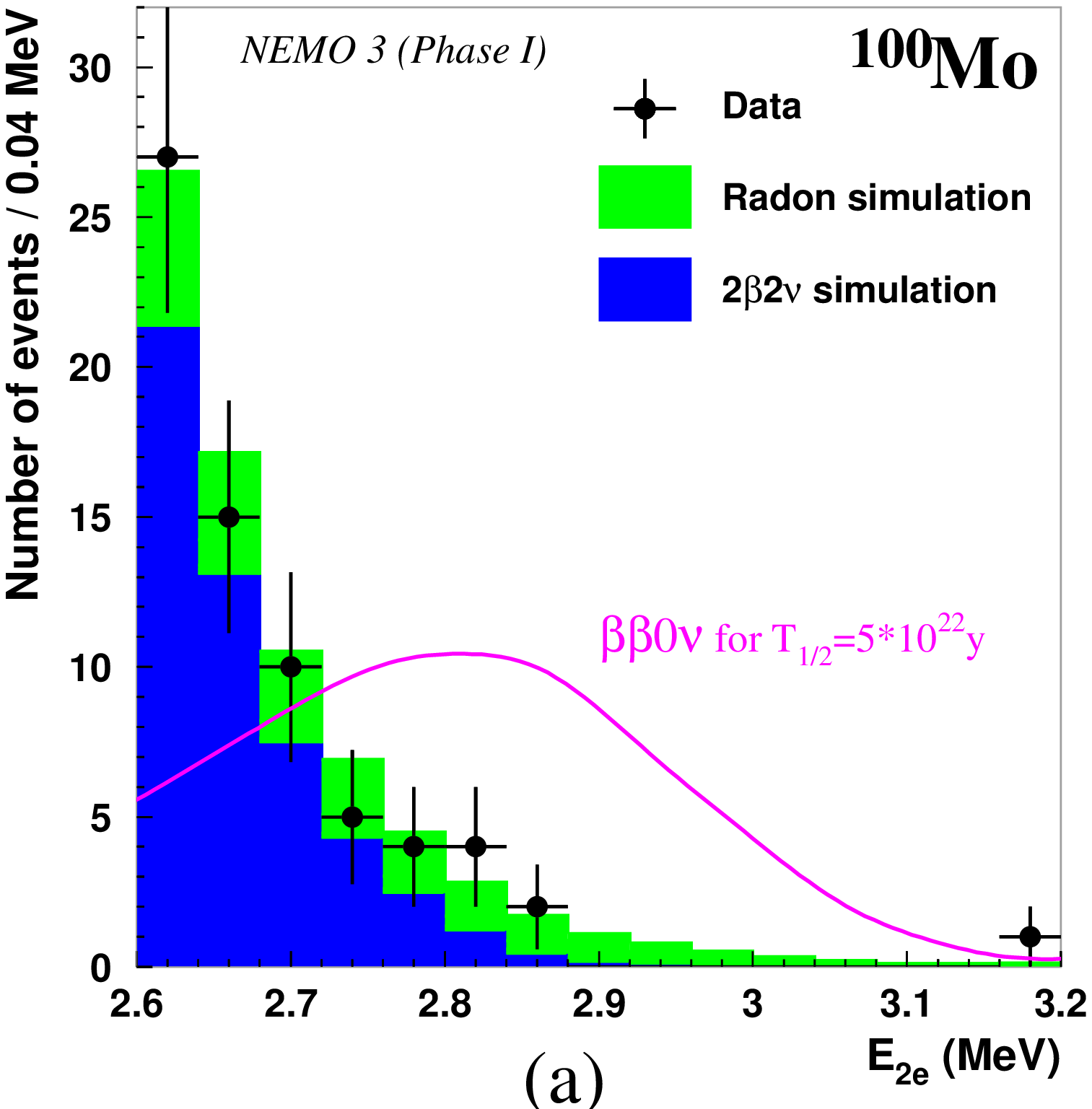}
\includegraphics[scale=0.3]{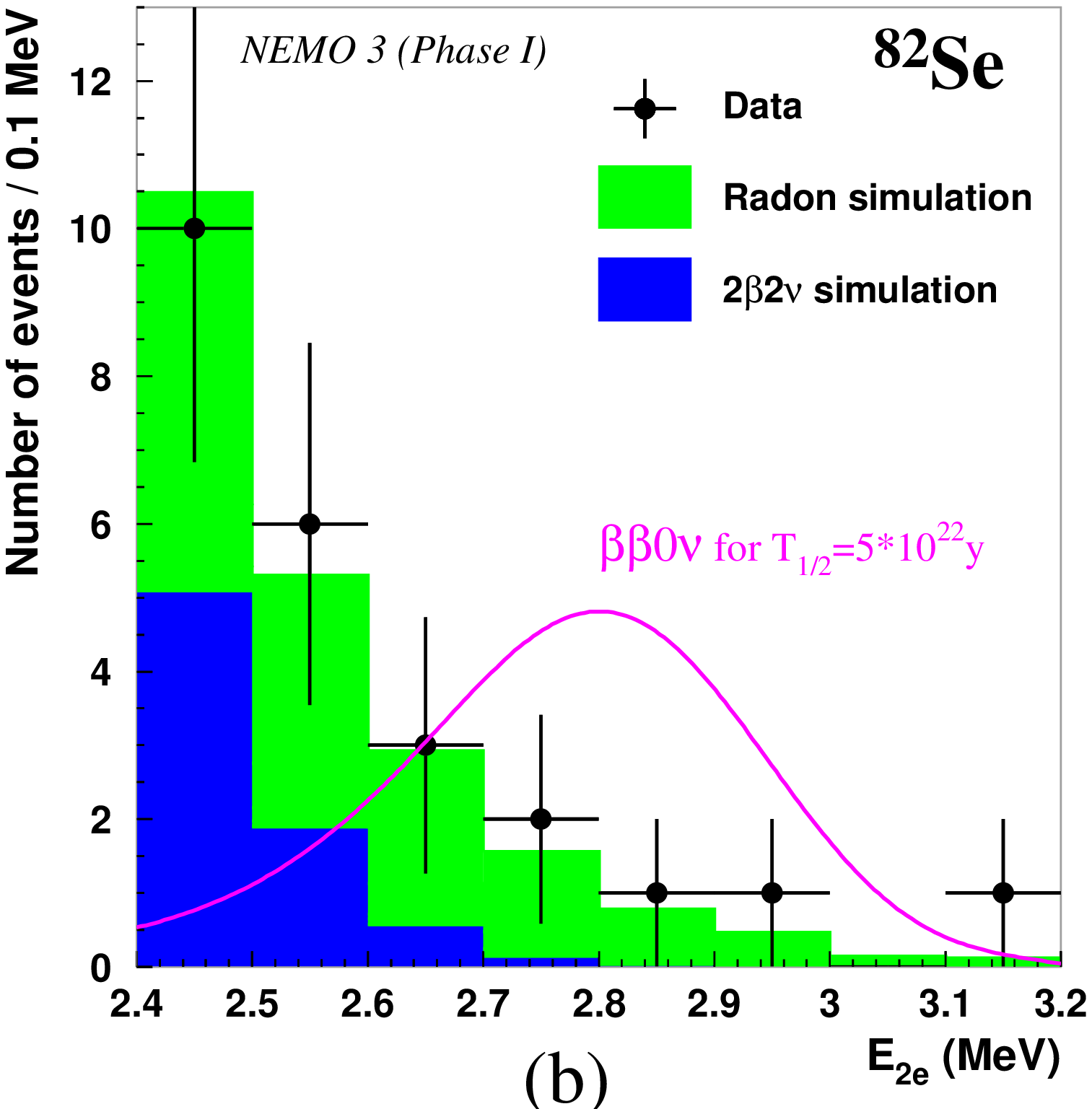}
\includegraphics[scale=0.3]{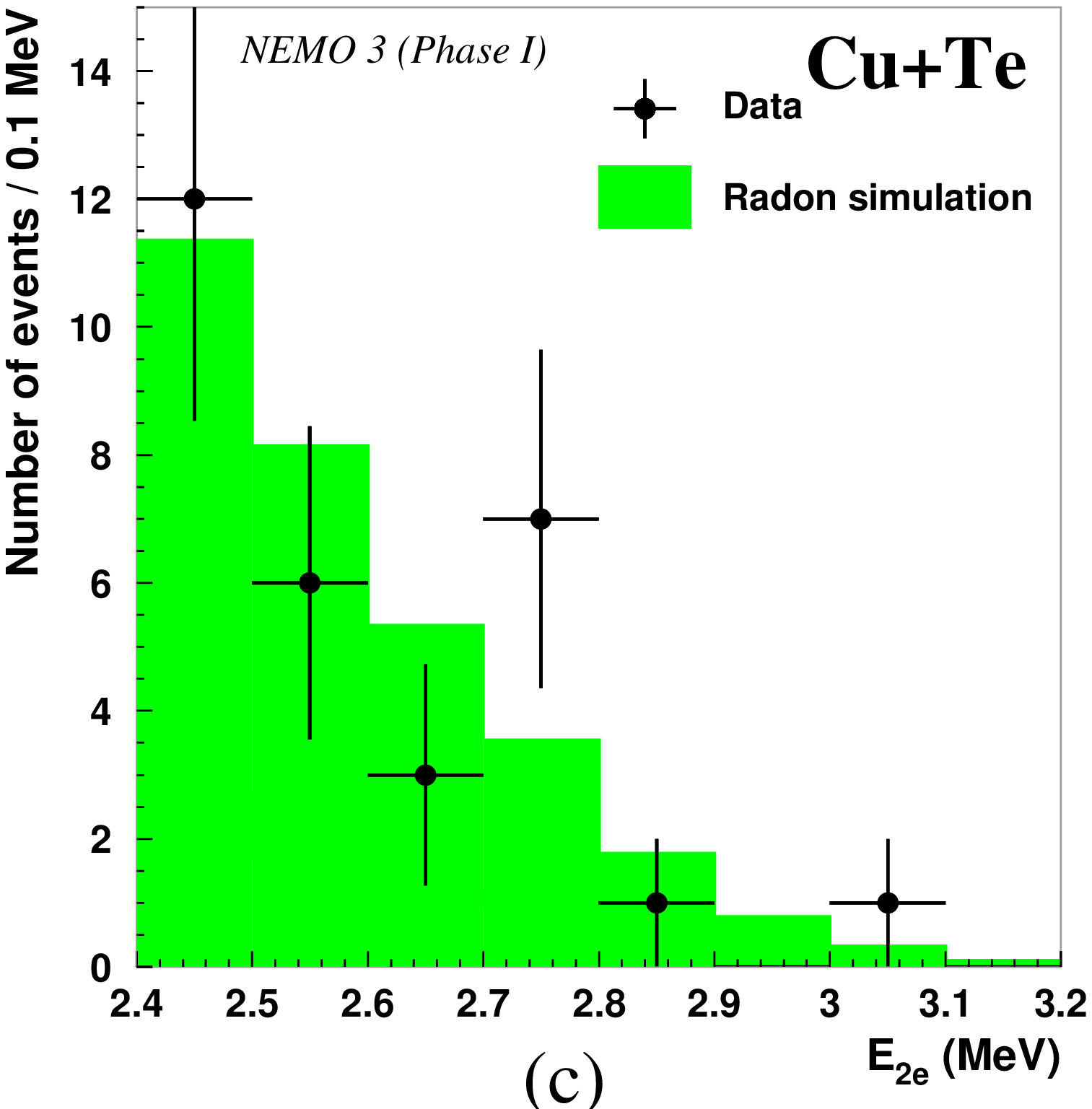}
\caption{\label{fig:results} Spectra of the energy sum of the two electrons in the $\beta\beta 0 \nu$ energy window 
after 389 effective days of data collection from February 2003 until September 2004 (Phase~I): 
(a) with  $6.914$~kg of $^{100}$Mo; (b) with $0.932$~kg of $^{82}$Se; (c) with Copper and Tellurium foils. 
The shaded histograms are the expected backrounds computed by Monte-Carlo simulations: 
dark (blue) is the $\beta\beta2\nu$ contribution and light (green) is the Radon contribution. 
The solid line corresponds to the expected $\beta\beta 0\nu$ signal if $T_{1/2}(\beta\beta0\nu)= 5 \times 10^{22}$ y.}
\end{figure*}

Fig.~\ref{fig:results}(a) and (b) show the tail of the two-electron energy sum spectrum in the 
$\beta\beta 0 \nu$ energy window for $^{100}$Mo and for $^{82}$Se respectively. 
The number of 2$e^-$  events observed in the data is in agreement with the expected number of 
events from $\beta\beta 2 \nu$ and the Radon simulations. 
For $^{100}$Mo, in the energy window $[2.8-3.2]$ MeV, the expected background is $8.1\pm1.3$ 
(error dominated by the uncertainty on the Radon activity) and 7 events have been observed. 
For $^{82}$Se, in the energy window $[2.7-3.2]$ MeV, the expected background is $3.1\pm0.6$ and 
5 events have been observed.
In order to independently check the dominant Radon contribution above 2.8 MeV, the energy sum spectrum 
(Fig.~\ref{fig:results}.c) has been plotted for the two electrons emitted from the Copper and Tellurium 
foils where no background except radon is expected. The data are in agreement with the Radon simulations.

In order to make the optimum use of all the information from the NEMO~3 detector, 
a maximum likelihood analysis \cite{Etienvre2003} has been applied to the 2$e^-$ event 
sample above 2 MeV using the three available variables: the energy sum ($E_{tot}$) of the two electrons, 
the  energy of each electron ($E_{min}$ is the minimum electron energy) and the angle between the two tracks ($cos\theta$). 
A three-dimensional probability distribution function, $P^{3D}$, can be written as:\\
\hspace*{3mm}$P^{3D}\, = \, P(E_{tot})\,P(E_{min}/E_{tot})\,P(cos\theta /E_{min})$\\
where $P(E_{min}/E_{tot})$ and $P(cos\theta/E_{min})$ are two conditional probability distribution functions. 
The likelihood is defined as\\
\hspace*{1cm}${\cal L} = \,  \prod_{i\, =\,1}^{N_{tot}} (\sum_{k=1}^8 \, x_k P^{3D}_k)$\\
where $k$ corresponds to one of the eight contributions: $\beta\beta 0 \nu$, $\beta\beta 2 \nu$, 
Radon, external and internal $^{214}$Bi and $^{208}$Tl, and neutrons. 
Here $x_k$ is the ratio of the number of 2$e^-$ events due to the 
process $k$ relative to the total number of observed events $N_{tot}$. 
Finally $P^{3D}_k$ is built using simulated events of contribution $k$. 

With 389 effective days of data collection, limits at 90\% C.L. obtained with the likelihood 
analysis are $T_{1/2}(\beta\beta0\nu)> 4.6 \times 10^{23}$~years  for $^{100}$Mo and 
$1.0 \times 10^{23}$~years for  $^{82}$Se. 
These limits are about 10 times higher than the previous 
limits obtained with $^{100}$Mo and $^{82}$Se \cite{Ejiri2001,Arnold1998}.
The corresponding upper limits for the effective Majorana neutrino mass range 
from 0.7 to 2.8 eV for $^{100}$Mo and 1.7 to 4.9 eV for $^{82}$Se depending on the nuclear matrix 
element calculation \cite{Caurier1996,Rodin2005,Simkovic1999,Suhonen1998,Suhonen2003,Stoica2001}. 
Results for each calculation are given in Table~\ref{tab:table1}.
For $^{100}$Mo,
since an incorrect value of the phase-space factor has been used in reference \cite{Stoica2001}, 
the value calculated in reference \cite{Vogel1992} has been used. 
The claim of a positive $\beta\beta 0 \nu$ signal observed with $^{76}$Ge \cite{Klapdor2004} 
gives an allowed effective mass range $0.1-0.9$~eV. Our limit obtained with 
$^{100}$Mo slightly overlaps this range. 
In the hypothesis of gluino or neutralino exchange, and using the nuclear matrix 
elements calculated in reference \cite{FaesslerL111}, limits obtained on the trilinear 
R-parity-violating supersymmetric coupling are $\lambda^{'}_{111} < 1.6 \times 10^{-4}$ 
for $^{100}$Mo and $3.0 \times 10^{-4}$ for $^{82}$Se.
In the hypothesis of a right-handed weak current, the limits are 
$T_{1/2}(\beta\beta0\nu)> 1.7 \times 10^{23}$~years at 90\% C.L. for $^{100}$Mo and 
$0.7 \times 10^{23}$~years for  $^{82}$Se, corresponding to an upper limit on the coupling 
constant of $\lambda  < 2.5 \times 10^{-6}$ for $^{100}$Mo and 
$3.8 \times 10^{-6}$ for $^{82}$Se using the nuclear calculations from references~\cite{Suhonen1998,Suhonen2002}.\\

\begin{table}
\caption{\label{tab:table1} Limits (in eV) on the effective neutrino mass $\langle m_{\nu} \rangle$ obtained from different theoretical calculations of nuclear matrix elements with $T_{1/2}(\beta\beta0\nu)>4.6 \times 10^{23}$y for $^{100}$Mo and $T_{1/2}(\beta\beta0\nu)>1.0 \times 10^{23}$y for $^{82}$Se.}
\begin{ruledtabular}
\begin{tabular}{cccc}
\multicolumn{2}{c}{Nuclear matrix elements} & $^{100}$Mo & $^{82}$Se\\
\hline
Shell model & Caurier 1996 \cite{Caurier1996}                  &                & $<$ 4.9       \\
QRPA & Rodin 2005 \cite{Rodin2005}                      & $<$ 2.7 - 2.8  & $<$ 4.1 - 4.5 \\
QRPA & Simkovic 1999 \cite{Simkovic1999}                & $<$ 1.0        & $<$ 3.3       \\
QRPA & Suhonen 2003 \cite{Suhonen1998,Suhonen2003}      & $<$ 1.1        & $<$ 2.8 - 4.2 \\
QRPA & Stoica 2001 \cite{Stoica2001}                    & $<$ 0.7 - 1.1  & $<$ 1.7 - 3.7 \\
\end{tabular}
\end{ruledtabular}
\end{table}

\section{Conclusions}

In conclusion, the NEMO~3 detector has been running reliably since February 2003. 
The $\beta\beta 2 \nu$ decay has been measured for $^{82}$Se and $^{100}$Mo with  
very high statistics and better precision than the previous measurements. The two-electron 
energy sum spectrum, the single energy spectrum and the angular distibution are all in good 
agreement with the $\beta\beta 2 \nu$ simulations. 
All components of the background in the $\beta\beta 0 \nu$ energy window have been 
measured directly using different analysis channels in the data. 
After 389 effective days of data collection, no evidence for $\beta\beta 0 \nu$ decay has been 
found in $^{100}$Mo or $^{82}$Se. The limits at the 90\% C.L. are 
$T_{1/2}(\beta\beta0\nu)>4.6 \times 10^{23}$~y for $^{100}$Mo and $1.0 \times 10^{23}$~y for  $^{82}$Se.
For this first running period (Phase~I) presented here, Radon was the dominant background at a level of
about 3 times higher than the $\beta \beta 2 \nu$ background for $^{100}$Mo. 
It has now been significantly reduced by a factor $\sim$10 by a radon-tight tent enclosing 
the detector and a radon-trap facility in operation since December 2004 which has 
started a second running period (Phase~II).
After five years of data collection, the expected sensitivity at 90\% C.L will be 
$T_{1/2}(\beta\beta0\nu) > 2 \times 10^{24}$~y for $^{100}$Mo and 
$8 \times 10^{23}$~y for $^{82}$Se, corresponding to 
$\langle m_{\nu} \rangle < 0.3-1.3$~eV for $^{100}$Mo and 
$\langle m_{\nu} \rangle < 0.6-1.7$~eV for $^{82}$Se.

A portion of this work was supported by grants from INTAS (03051-3431, 03-55-1689) and NATO (PST CLG 980022)


\begin{thebibliography}{6}
\bibitem{SuperKSolar} Y. Fukuda et al., Phys. Rev. Lett. 86 (2001) 5651
\bibitem{SuperKAtm} Y. Fukuda et al., Phys. Rev. Lett. 81 (1998) 1562
\bibitem{SNO} Q.R. Ahmad et al., Phys. Rev. Lett. 89 (2002) 011301
\bibitem{Kamland} T. Araki et al., Phys. Rev. Lett. 94 (2005) 081801
\bibitem{Mainz} C. Weinheimer et al., Nucl. Phys. B 118 (2003) 279
\bibitem{Troitsk} V. Lobashev et al.,  Nucl. Phys. B  91 (2001) 280
\bibitem{Mohapatra1980} R.N. Mohapatra et al., Phys. Rev. Lett. 44 (1980) 912
\bibitem{Leptogenesis} M. Fukugita et al., Phys. Lett. B174 (1986) 45
\bibitem{Augier2005} R. Arnold et al., Nucl. Inst. Meth. A536 (2005) 79
\bibitem{PDG}Particle Data Group, K. Hagiwara et al., Phys. Rev. D66 (2002) 010001
\bibitem{Etienvre2003} A.I. Etienvre, PhD Thesis, University Paris-Sud (2003)
\bibitem{Takeuchi1999} Y. Takeuchi et al., Nucl. Inst. Meth. A421 (1999) 334
\bibitem{Ejiri2001} H. Ejiri et al., Phys. Rev. C63 (2001) 065501
\bibitem{Arnold1998} R. Arnold et al., Nucl. Phys. A636 (1998) 209
\bibitem{Caurier1996} E. Caurier et al., Phys. Rev. Lett. 77 (1996) 1954
\bibitem{Rodin2005} V.A. Rodin et al., Phys. Rev. C68 (2003) 044302, nucl-th/0503063
\bibitem{Simkovic1999} F. Simkovic et al., Phys. Rev. C60 (1999) 055502
\bibitem{Suhonen1998} M. Aunola et al., Nucl. Phys. A643 (1998) 207
\bibitem{Suhonen2003} J. Suhonen et al., Nucl. Phys. A723 (2003) 271
\bibitem{Stoica2001} S. Stoica et al., Nucl. Phys. A694 (2001) 269
\bibitem{Vogel1992} F. Boehm and P. Vogel, Physics of massive neutrinos, Cambridge Univ. Press, second edition (1992)
\bibitem{FaesslerL111} A. Faessler et al., Phys. Rev. D58 (1998) 115004
\bibitem{Suhonen2002} J. Suhonen, Nucl. Phys. A700 (2002) 649
\bibitem{Klapdor2004} H.V. Klapdor-Kleingrothaus et al.,  Nucl. Inst. Meth. A522 (2004) 371
\end{thebibliography}
\end{document}